\documentclass[twocol]{ametsocV6.1}

\usepackage{amsmath,amsfonts,amssymb,bm}
\usepackage{mathptmx}
\usepackage{newtxtext}
\usepackage{newtxmath}
\usepackage{subcaption}
\usepackage[normalem]{ulem}

\ifdraft\else
\PassOptionsToPackage{hyphens}{url}
\usepackage{hyperref}
\hypersetup{
	breaklinks,
	colorlinks=true,
	linkcolor=blue,
    urlcolor=Purple,
	citecolor=Purple,
 }
\fi

\newcommand{\Ri}{\mathrm{Ri}}
\newcommand{\bu}{\boldsymbol{u}}

\title{Surface heating steers planetary-scale ocean circulation}

\authors{Dhruv Bhagtani,\aff{a}\correspondingauthor{D.~Bhagtani, dhruv.bhagtani@anu.edu.au} 
Andrew McC. Hogg,\aff{a} 
Ryan M. Holmes,\aff{b} and
Navid C. Constantinou\aff{a}
}

\affiliation{\aff{a}{Research School of Earth Sciences \& ARC Center of Excellence for Climate Extremes,\ifdraft\else \\ \fi Australian National University, Canberra, Australia}\\
\aff{b}{School of Geosciences, University of Sydney, Sydney, Australia}}

\abstract{
Gyres are central features of large-scale ocean circulation and are involved in transporting tracers such as heat, nutrients, and carbon-dioxide within and across ocean basins. Traditionally, the gyre circulation is thought to be driven by surface winds and quantified via Sverdrup balance, but it has been proposed that surface buoyancy fluxes may also contribute to gyre forcing. Through a series of eddy-permitting global ocean model simulations with perturbed surface forcing, the relative contribution of wind stress and surface heat flux forcing to the large-scale ocean circulation is investigated, focusing on the subtropical gyres. In addition to gyre strength being linearly proportional to wind stress, it is shown that the gyre circulation is strongly impacted by variations in the surface heat flux (specifically, its meridional gradient) through a rearrangement of the ocean's buoyancy structure. On shorter timescales ($\sim$ decade), the gyre circulation anomalies are proportional to the magnitude of the surface heat flux gradient perturbation, with up to $\sim 0.15\,\mathrm{Sv}$ anomaly induced per $\mathrm{W}\,\mathrm{m}^{-2}$ change in the surface heat flux. On timescales longer than a decade, the gyre response to surface buoyancy flux gradient perturbations becomes non-linear as ocean circulation anomalies feed back onto the buoyancy structure induced by the surface buoyancy fluxes. These interactions complicate the development of a buoyancy-driven theory for the gyres to complement the Sverdrup relation. The flux-forced simulations underscore the importance of surface buoyancy forcing in steering the large-scale ocean circulation. 
}
\begin{document}

\maketitle

\maketitle


\statement{
Ocean gyres are large swirling circulation features that redistribute heat across ocean basins. It is commonly believed that surface winds are the sole driver of ocean gyres, but recent literature suggests that other mechanisms could also be influential. We perform a series of numerical simulations in which we artificially change either the winds or the heating at the ocean's surface and investigate how each factor independently affects the ocean gyres. We find that gyres are steered by both winds and surface heating, and that the ocean circulation responds differently to heating on short and long timescales. In addition, the circulation depends on where the heating is applied at the ocean's surface. Through these simulations, we argue that a complete theory about ocean gyres must consider heating at the ocean's surface as a possible driver, in addition to the winds.
}
%
\section{Introduction}

The large-scale ocean circulation derives its energy from a variety of sources including wind stress \citep{Wunsch2004, Hughes2008, Jamet2021}, tidal forces \citep{Oka2013}, and surface and geothermal buoyancy fluxes \citep{Hughes2009, Hogg2020}. Gyres are fundamental elements of the large-scale circulation and play a crucial role in the biogeochemical and hydrological cycles in the ocean by transporting momentum, heat, nutrients and chemicals within and across ocean basins \citep{Webb2012}. In particular, gyres contribute to global heat transport by transferring heat poleward \citep{Palter2015, Zhang2021, Li2022}. Despite their importance in regulating large-scale weather and climate patterns, the interplay between the processes leading to the formation and evolution of ocean gyres are not fully understood.

Traditional oceanographic literature on the drivers of ocean circulation suggest that horizontal flows are primarily caused by mechanical forcing due to wind stress \citep{Sverdrup1947} and tidal forces \citep{Wunsch2004, Oka2013}, and the meridional overturning circulation (MOC) is chiefly driven by surface buoyancy fluxes \citep{Stommel1959}. However, this simplified viewpoint has been amended with time. Recent literature points to a definitive role played by wind stress on the MOC through isopycnal upwelling in the Southern Ocean \citep{Abernathey2015, Hogg2017} and bottom-enhanced diapycnal mixing \citep{Stanley2014, Drake2020}. Similarly, buoyancy forcing is thought to exert a significant control on horizontal circulation through the conversion from potential to kinetic energy \citep{Tailleux2009, Hughes2009} and control of the stratification in the ocean \citep{Shi2020}, consistent with the fundamental dynamics of rotating horizontal convection \citep{Gayen2021}. In this paper, we aim to evaluate the interplay between wind stress and surface buoyancy forcing in driving various features of the large-scale ocean circulation, with an emphasis on basin-scale ocean gyres.

Considerable progress has been made to elucidate the impact of surface wind stress on ocean gyres \citep{Sverdrup1947, Stommel1948, Munk1950, Rhines1982, Luyten1983, Pedlosky1986}. \citet{Sverdrup1947} proposed what became a namesake relationship that links the curl of the wind stress and the depth-integrated meridional geostrophic transport, valid in the ocean interior away from coastlines,
\begin{equation}
    V = \frac{\hat{\boldsymbol{z}} \boldsymbol{\cdot} (\boldsymbol{\nabla} \times \boldsymbol{\tau})}{\beta},
    \label{eq:Sverdrup}
\end{equation}
where $V = \rho_0 \int \upsilon \,\mathrm{d}z$ is the time-mean depth-integrated meridional mass transport with $\rho_0$ the ocean's reference density, $\hat{\boldsymbol{z}}$ is the unit vector in the vertical, $\boldsymbol{\tau} = \tau_x \, \hat{\boldsymbol{x}} + \tau_y \, \hat{\boldsymbol{y}}$ the time-mean horizontal wind stress at the ocean's surface, and $\beta = \partial f / \partial y$ is the meridional gradient of the Coriolis frequency $f$. The return flow occurs as an intense inertial western boundary flow (see, e.g., \cite{Hughes2001}). The Sverdrup relation~\eqref{eq:Sverdrup} describes the dependence of the horizontal structure of barotropic gyres on the wind stress curl, and to date, remains the corner-stone theory of wind-driven gyres.

The Sverdrup relation focuses on understanding the horizontal structure of the vertically-integrated gyre transport. The ventilated thermocline theory, developed by \citet{Luyten1983}, made further strides in interpreting the vertical structure of ocean gyres forced by wind stress. They obtained a layer-wise meridional transport for gyres but restricted gyre transport to only ventilated isopycnals; that is, isopycnals outcropping to the surface of the ocean, with non-ventilated isopycnals at rest. However, several studies point to a net re-circulatory transport in the non-ventilated isopycnal layers \citep{Rhines1982, McDowell1982} so long as they do not interact with the ocean's surface or topography. Therefore, ocean gyres can be viewed as a combination of ventilation and recirculation regimes, and are controlled by surface wind stress curl.

Wind-driven theories encapsulate gyre circulation to zeroth order, however, observational studies show deviations from the Sverdrup relation. \citet{Gray2014} examined the validity of Sverdrup dynamics in a point-wise manner using observations from Argo floats \citep{Roemmich2004} and found that it agrees well with observations in the interior subtropical gyres, with significant deviations in subpolar regions. \citet{Verdiere2016} using data from Argo \citep{Ollitrault2013} and the World Ocean Atlas \citep{Locarnini2010} computed the global Sverdrup streamfunction and found that the strength of the gyres was underrepresented by a factor of 2. These discrepancies indicate that processes other than Sverdrup dynamics could be at play in establishing the gyre strength, such as bottom pressure torques \citep{Hughes2001}, diapycnal mixing \citep{Lavergne2022}, buoyancy forcing \citep{Hogg2020, Liu2022}, and coupling with the meridional overturning \citep{Klockmann2020, Berglund2022}. A unified theory encapsulating all the previously stated mechanisms is lacking and this limits our understanding of the coupling between these processes, as well as their combined effects on ocean circulation. The present paper makes an attempt to isolate and understand the roles of wind stress and surface buoyancy forcing in shaping the planetary-scale ocean gyres.

Surface buoyancy forcing alters the ocean's density structure through heat and freshwater fluxes \citep{Large2009, Talley2011}. Together with the thermal wind relation,
\begin{equation}
    f \frac{\partial \boldsymbol{u}}{\partial z} = \boldsymbol{\hat{z}} \times \boldsymbol{\nabla} b,
    \label{eq:TWR}
\end{equation}
where $b(\boldsymbol{x}, t) = g (1 - \rho(\boldsymbol{x}, t) / \rho_0)$ is the buoyancy, with $g$ the gravitational acceleration, these density structure changes can be used to understand how changes in surface buoyancy forcing might impact ocean circulation. However, the relationship between surface buoyancy forcing and the horizontal circulation is complicated. For example, variations in mixed layer depth and non-linear feedbacks with the ocean circulation both influence how the surface buoyancy forcing is ``felt'' within the ocean. The mixed layer ingests a fraction of the surface buoyancy flux, which is reflected in the anomalous ocean's density within the layer. The amount of buoyancy forcing reaching the layers below is thus inversely related to the mixed layer depth, with a deeper mixed layer taking a longer time to relay the excess buoyancy forcing into the subsurface layers \citep{Xie2010} due to its higher effective heat capacity. The effect of surface heat flux on the near-surface buoyancy also depends on the thermal expansion coefficient of seawater, which varies spatially by an order of magnitude across the globe and is important in setting the ocean's stratification \citep{Caneill2022}. Furthermore, the circulation modifies the influence of surface buoyancy forcing on the ocean's buoyancy structure through heat advection \citep{Bryden1991}. Advection acts to alter the buoyancy structure remote from the forcing, which, in view of~\eqref{eq:TWR}, would also cause anomalies in the ocean circulation in that remote location. In this paper, we evaluate the variability in ocean's stratification over time to better understand the non-linear and non-local connection between the surface buoyancy forcing and the gyres.

Past studies examined the role of surface buoyancy forcing in restructuring ocean gyres. \citet{Goldsbrough1933} observed that freshwater fluxes can drive horizontal circulation via induced sea surface height anomalies. \citet{Luyten1985} and \citet{Pedlosky1986} extended the ventilated thermocline theory \citep{Luyten1983} to include an interfacial mass flux between various isopycnal layers (to represent surface buoyancy forcing), and using a simplified ocean model, demonstrated a geostrophic baroclinic flow induced by the buoyancy flux and steered by wind stress. \citet{Verdiere1989} coupled the surface buoyancy forcing to wind stress via a bulk formula and concluded that the former drives a baroclinic mode to significantly recast the horizontal and vertical structure of subtropical gyres. Model studies that forced the ocean using only surface buoyancy fluxes observed single or double gyre circulation; see for example \citet{Winton1997, LaCasce2004, Nilsson2005}. \citet{Gjermundsen2018} also conducted simulations forced only via a meridionally-varying surface temperature restoring profile and observed a broad eastward zonal flow and a western boundary current. \citet{Hogg2020} conducted a series of numerical simulations with a restoring temperature profile and no wind stress in two configurations: a direct numerical simulation in a 3D box domain, and a layered general circulation model in a sector configuration, and found a double (resembling a subtropical and subpolar) gyre in both scenarios. With a multitude of contrasting viewpoints on the processes leading to the formation of gyres, we are motivated to examine further the role of surface buoyancy forcing in driving ocean gyres.

It is worth emphasizing here that ocean gyres do not exist in isolation -- they interact with other fundamental aspects of ocean circulation, such as the MOC. The MOC complements gyre circulation in transporting tracers across ocean basins. It has been argued that thermohaline forcing is responsible for driving the MOC \citep{Stommel1959, Stommel1961}, with gyres being driven by wind stress \citep{Sverdrup1947}. \citet{Luyten1985} contradicted this simplified view by establishing a link between wind stress and heat gain for the North Atlantic subpolar gyre. \citet{Yeager2014} and \citet{Yeager2015} used a coupled ocean--sea ice configuration of the Community Earth System Model with perturbed forcing, and found that most of the decadal variability in both the Atlantic MOC (AMOC) and the North Atlantic subpolar gyre was due to variability in surface buoyancy forcing, while interannual variability in these circulatory features were attributed to wind stress anomalies. Modeling studies \citep{Yeager2015, Larson2020} identify a direct relationship between the mid-depth overturning cell and the North Atlantic subtropical gyre transport on the basis that the two circulatory features are linked through the northward flowing Gulf Stream. Thus, ocean gyres and the MOC are coupled dynamical features \citep{Klockmann2020, Berglund2022}, and a thorough analysis of the drivers behind the formation of ocean gyres requires a quantitative understanding of other processes of ocean circulation.

The objectives of our paper are: \emph{(i)}~to quantify how the surface buoyancy forcing affects the structure of ocean gyres, and \emph{(ii)}~to understand how wind stress and surface buoyancy fluxes act in concert to drive large-scale ocean circulation. Section~\ref{Models_methods} outlines the simulation setup and a gyre metric used to analyze the ocean's circulation. Section~\ref{Results_wind} examines the sensitivity of gyre circulation to changes in surface wind stress, followed by a brief discussion of the coupling between gyres and other large-scale circulation features in the ocean. Section~\ref{Results_buoy_contrast} further investigates the role of surface buoyancy forcing gradients in steering the ocean circulation. In section~\ref{sec:Global_warming} we look at a uniform warming experiment to illustrate that changes in ocean circulation can also occur in the absence of a meridional gradient in buoyancy forcing anomaly. In section~\ref{Conclusions}, we conclude by emphasizing the connected roles of wind stress and surface buoyancy fluxes in driving ocean circulation, the importance of surface buoyancy forcing in driving ocean gyres, and future directions to advance our understanding of ocean circulation.

\section{Models and Methods}
\label{Models_methods}
\subsection{Flux-forced simulations}

Surface buoyancy fluxes in ocean--sea ice general circulation models are usually parameterized using bulk formulae \citep{Large1994}, and are therefore dependent on the model's dynamic sea surface temperature, as well as the externally prescribed atmospheric winds, humidity, air temperature and radiative fluxes. Therefore, any changes in circulation (for example due to changes in wind stress) have the ability to alter the surface buoyancy forcing. In this study we construct a series of global simulations in which we force the ocean using prescribed fluxes at the surface, allowing us to independently modify the surface boundary fluxes. We call them ``flux-forced'' simulations.

Forcing for the flux-forced control experiment is constructed from a 200-year control simulation using ACCESS-OM2-025 \citep{Kiss2020}, a global ocean--sea ice model at 0.25$^\circ$ resolution and 50 vertical layers. ACCESS-OM2-025 is an amalgamation of the Modular Ocean Model~v5.1 \citep{Griffies2012} and the CICE~v5.1.2 \citep{Hunke2015} sea ice model. We drive the ACCESS-OM2-025 control simulation using a repeat-year atmospheric forcing from the JRA55-do~v1.3 reanalysis product \citep{Tsujino2018}. We use the period 1st May 1990 to 30th April 1991 as the repeat year for atmospheric forcing following \citet{Stewart2020}. The ACCESS-OM2-025 control experiment is initialized using temperature and salinity data from the World Ocean Atlas 2013 \citep{Locarnini2013, Zweng2013}, and incorporates the Gent-McWilliams parameterization \citep{Gent1990} (with a variable diffusivity, limited to $200\; \mathrm{m}^2\,\mathrm{s}^{-1}$) to complement partially resolved mesoscale eddy fluxes. Vertical mixing is parameterized using a modified K-profile parameterization \citep{Large1994} (see Appendix for details on these modifications). We use the last 20 years from the 200-year ACCESS-OM2-025 control simulation to create a climatology of surface boundary fluxes at 3-hourly temporal frequency with which we force the flux-forced simulations.

\begin{figure*}[t]
    \noindent
    \includegraphics[width=\textwidth]{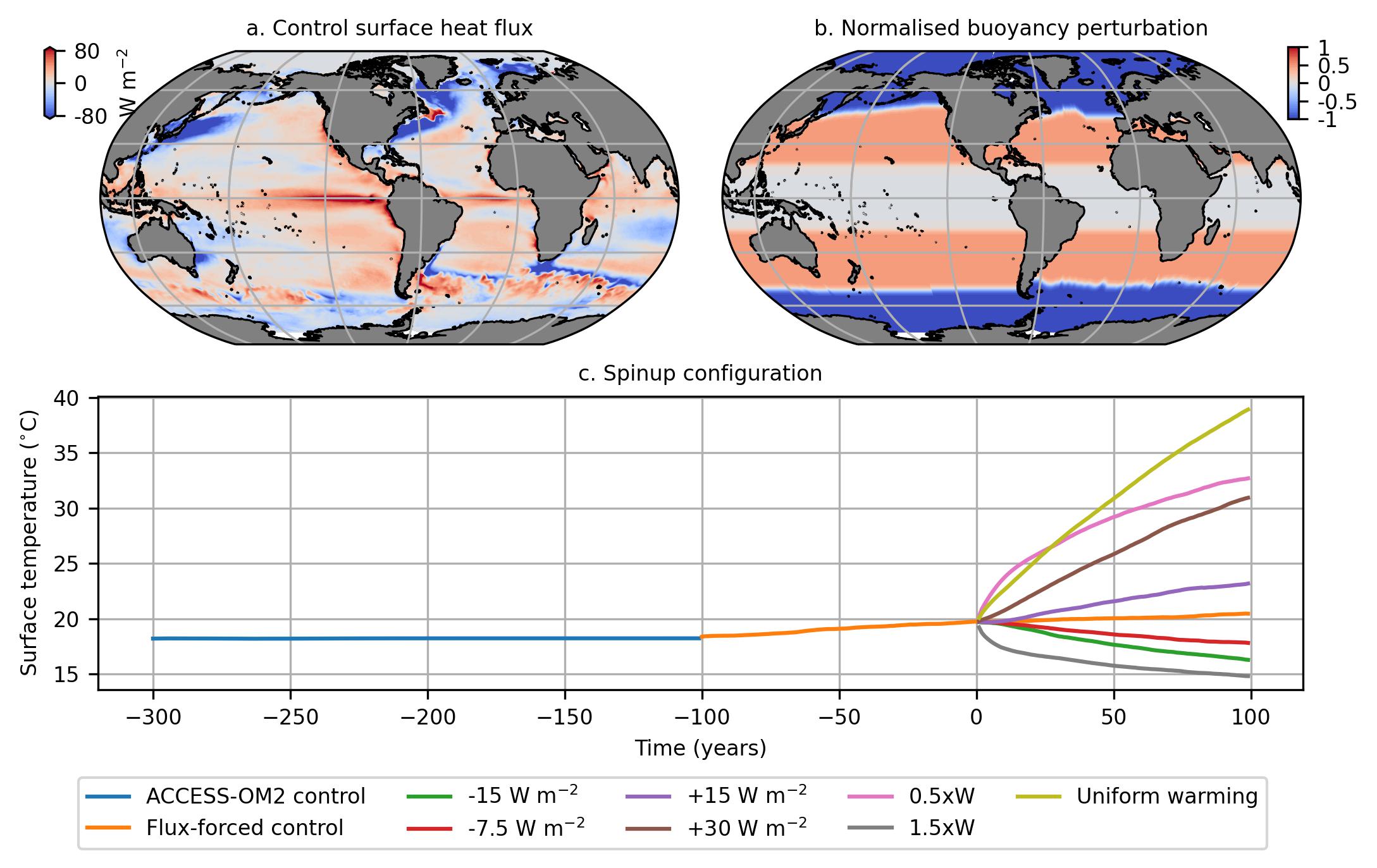}
    \caption{Model setup for sensitivity experiments. (a)~Climatological net surface heating for flux-forced control simulation. (b)~Surface buoyancy flux perturbation pattern. This pattern is multiplied by a scale factor and then applied to panel~(a) to construct the surface buoyancy flux perturbation experiments. The perturbation pattern is $-1$ in subpolar and polar regions, and $+0.5$ in subtropical regions; a $1.5$ meridional contrast between the two extrema. A hyperbolic tangent function smoothly connects the: \emph{(i)}~subtropical and subpolar regions and \emph{(ii)}~subtropics and tropics. (c)~Global average surface temperature from each of our simulations, illustrating the model spinup method. Simulation time is referenced with respect to the beginning of the flux-forced perturbation experiments.}
    \label{fig:expt_config}
\end{figure*}

The 3-hourly climatology of surface boundary fluxes is used to force a stand-alone implementation of the Modular Ocean Model~v5.1. The flux-forced control experiment is initialized from the end of the ACCESS-OM2 control experiment and run for 100 years, after which we branch off a series of flux-forced perturbation simulations with modified surface fluxes. These perturbation experiments are run for another $100$ years. 
Although 100 years is not sufficient for the deep ocean to reach equilibrium, it is enough for the upper and mid-ocean circulation to respond to changes in surface forcing \citep{Saenko2009}. 

We conduct three types of sensitivity experiments:
\begin{itemize}
    \item[\emph{(i)}]
     Perturbations in surface wind stress,
    \item[\emph{(ii)}]
     Perturbations in surface meridional heat flux gradients, and
    \item[\emph{(iii)}]
     A `uniform warming' perturbation.
\end{itemize}
A list of all flux-forced experiments is given in Table~\ref{T1:model_setup}. Each set of experiments is described below.

\begin{table*}[h]
\caption{List of flux-forced experiments. G denotes perturbation applied globally and G$-$T perturbations that exclude the tropics, that is the equatorward extent of subtropical gyres in the equilibrated flux-forced control simulation (Fig.~\ref{fig:expt_config}b).}\label{T1:model_setup}
\begin{center}
\begin{tabular}{ccccrrcrc}
\topline
Experiment & Wind Factor & Surface Buoyancy Flux Contrast $\Delta B$ ($\mathrm{W}\,\mathrm{m}^{-2}$) & Region\\
\midline
 Control & 1 & 0 & G \\
 0.5$\times$W & 0.5 & 0 & G \\
 1.5$\times$W & 1.5 & 0 & G \\
 $-15 \;\mathrm{W}\,\mathrm{m}^{-2}$ & 1 & $-15$ & G$-$T \\
 $-7.5\;\mathrm{W}\,\mathrm{m}^{-2}$ & 1 & $-7.5$ & G$-$T \\
 $+15 \;\mathrm{W}\,\mathrm{m}^{-2}$ & 1 & $+15$ & G$-$T \\
 $+30 \;\mathrm{W}\,\mathrm{m}^{-2}$ & 1 & $+30$ & G$-$T \\
 Uniform warming & 1 & 0, instead \emph{spatially uniform} +5 & G \\
\botline
\end{tabular}
\end{center}
\end{table*}

We perform wind perturbation experiments by increasing or decreasing the global wind stresses by a multiplicative factor of 0.5 or 1.5 respectively (Table~\ref{T1:model_setup}). 

The construction of surface buoyancy flux gradient perturbation experiments is based on the thermal wind relation~\eqref{eq:TWR}, which suggests a dependence of the ocean circulation on horizontal density gradients, which in turn could be modified by prescribing a spatially varying buoyancy flux perturbation at the ocean's surface. Herein, we apply buoyancy flux perturbations by varying the prescribed surface heat fluxes. The largest heat losses in the ocean occur in focused regions over the subtropical western boundary currents (Fig.~\ref{fig:expt_config}a). The fine-scale spatial structure in the surface heat fluxes differs from the broader patterns of the wind stress forcing, and for that reason, choosing a multiplicative approach for surface buoyancy flux perturbations would enhance this fine-scale structure, which could instigate spurious behavior in ocean circulation. Furthermore, a multiplicative approach may lead to strong convection over the western boundary currents where heat loss in the control is strongest (Fig.~\ref{fig:expt_config}a). Therefore, for our buoyancy flux gradient perturbations we choose instead to add or subtract a broad surface heat flux pattern (Fig.~\ref{T1:model_setup}b) to the flux-forced control simulation, multiplied by a constant flux amplitude, $\Delta B$, for each experiment (Table~\ref{T1:model_setup}) to enhance or reduce the meridional buoyancy gradients. 

The forcing pattern in Fig.~\ref{fig:expt_config}b is chosen to modify meridional density gradients at the northward and southward edges of the subtropical gyres and their impact on gyre transport (as suggested by~\eqref{eq:TWR}). We ensure that the global integral of the pattern is zero by adjusting the magnitude of the heat flux in subpolar and polar regions to be double the magnitude of heat flux in subtropical regions (which has twice the area), with zero perturbation in the tropics. 
To minimize any spurious behavior in ocean circulation due to the applied buoyancy perturbation, we employ a hyperbolic tangent function ($\tanh{[(y-y_0)/\Delta y]}$ over the latitude band of $\Delta y = 12.5^{\circ}$, with $y_0$ the transition latitude; see Fig.~\ref{fig:expt_config}b) at the junction between subtropical and subpolar buoyancy anomalies. Buoyancy perturbation experiments for which the mask is multiplied by a positive value have surface heat fluxes that enhance the near-surface meridional buoyancy gradients, and are labeled as ``increased buoyancy flux contrast'' experiments. Conversely, all experiments where the buoyancy perturbation mask is multiplied by a negative value are labeled as ``reduced buoyancy flux contrast'' experiments. We were unable to run a $-30 \;\mathrm{W}\,\mathrm{m}^{-2}$ simulation as it became unstable due to unrealistically warm sea surface temperatures at high latitudes.

Finally, we conduct a globally uniform warming experiment, which differs from the surface meridional heat flux gradient perturbation experiments in that we do not externally induce a surface buoyancy gradient in the ocean. However, we still anticipate anomalies in the circulation in this experiment owing to its non-local and non-linear advective feedbacks with the surface buoyancy forcing, and lateral variations in the mixed layer depth and the thermal expansion coefficient.

The flux-forced control simulation is not fully equilibrated as shown in Fig.~\ref{fig:expt_config}c, where the mean sea surface temperatures gradually increase with time ($\approx0.1^{\circ}\,\mathrm{C}\,\mathrm{decade}^{-1}$). This systematic increase is due to frazil formation in polar latitudes, which is modeled via an additional heat input. This heat gain is a proxy for heat transferred by a fictional ice model coupled to the flux-forced ocean model, as cold water is converted to ice. Frazil formation in our experiments is not prescribed like the other surface heat fluxes; instead, it depends on ocean temperature and acts to alleviate excessive cooling in polar regions. The dependence of frazil formation on the surface buoyancy forcing limits the magnitude of buoyancy flux perturbation we can apply at the ocean's surface without the simulation becoming unstable. Moreover, the heat gain due to frazil formation is amplified in increased buoyancy flux contrast experiments with a stronger heat loss in subpolar and polar regions. This net heat gain is partially mitigated by applying a globally uniform heat loss of $1.28\,\mathrm{W}\,\mathrm{m}^{-2}$ to all increased buoyancy flux contrast experiments. The resulting heat flux anomalies due to frazil formation are smaller than the buoyancy perturbations applied in our sensitivity experiments. The frazil offset implies that we need to apply larger surface buoyancy flux gradients in the increased buoyancy flux contrast experiments to observe significant circulation changes (see Table~\ref{T1:model_setup}). 

\subsection{Gyre metrics}

Gyre strength is generally defined as the vertically integrated mass transport from surface to bottom.
However, this procedure disregards the baroclinic component of gyre strength, which integrates to zero in depth. To circumvent this issue, we estimate the near-surface subtropical gyre strength using an ``isopycnal outcropping method". We integrate the meridional mass transport from the surface only to the depth of the densest isopycnal (measured using potential density referenced to $2000\;\mathrm{dbar}$ and denoted as $\sigma_{\max}$) that outcrops to the ocean's surface in a given basin (marked by the red boxes in Fig.~\ref{fig:wind_driven_gyres}). For simplicity, we choose $\sigma_{\max} = 1035.8\;\mathrm{kg}\,\mathrm{m}^{-3}$ for all four subtropical gyres in each flux-forced simulation (which roughly captures the top 800-1200 $\mathrm{m}$ of the ocean). The horizontal transport streamfunction for a basin is computed by cumulatively integrating the meridional mass transport in the zonal direction. Then, to arrive at a single scalar estimate of the gyre's strength, we select the 95$^\text{th}$ percentile (to filter out vigorous inertial re-circulating eddies near the western boundary region) of a 5-year running mean (to filter out transient eddies and seasonal isopycnal outcropping) of the resulting density-integrated horizontal transport streamfunction.

The isopycnal outcropping method captures the baroclinic component (or the gravest surface modes, as stated by \citet{Lacasce2020}) of gyres, and is used in all flux-forced simulations to compare the gyre strength. However, it suffers from two limitations:
\begin{enumerate}
    \item
    Surface buoyancy perturbations could restructure the ocean's stratification, which may alter the isopycnal regime occupied by the gyres. These changes are not well-represented in the isopycnal outcropping method, since the method integrates the entire circulation from the surface to $\sigma_{\max} = 1035.8\;\mathrm{kg}\,\mathrm{m}^{-3}$. We use a deep isopycnal for $\sigma_{\max}$ to ensure we fully capture the subtropical gyres. In doing so, we may record a portion of abyssal circulation, which is usually much weaker than near-surface circulation. Thus, this method characterizes the subtropical gyre strength.
    \item
    Computing the streamfunction requires that the flow is divergence-free, which is not guaranteed due to the possibility of a net transport across the $\sigma_{\max}$ isopycnal. However, in the ocean's interior, flow across isopycnals is weak compared with flow along isopycnals \citep[see, e.g.,][]{Abernathey-etal-2020-oceanmixing} and, therefore, our streamfunction calculations are correct to leading order.
\end{enumerate}

\section{Wind stress perturbation experiments}
\label{Results_wind}

In this section, we investigate two perturbation experiments wherein we change the magnitude of wind stress by 0.5 and 1.5 times the control experiment, which alters the time-mean vorticity input due to the wind stress curl by the same factor. We analyze short-term and long-term variations in the subtropical gyre transport for Atlantic and Pacific Oceans, along with a brief discussion of their coupling with the AMOC. Since our flux-forced experiments do not incorporate sea-ice dynamics, changes in Weddell and Ross gyres are not reported in the paper, as sea-ice can significantly alter gyre dynamics in polar regions.

\begin{figure*}[t]
    \centering
    \includegraphics[width=\textwidth, trim={16cm 0cm 10cm 0cm}]{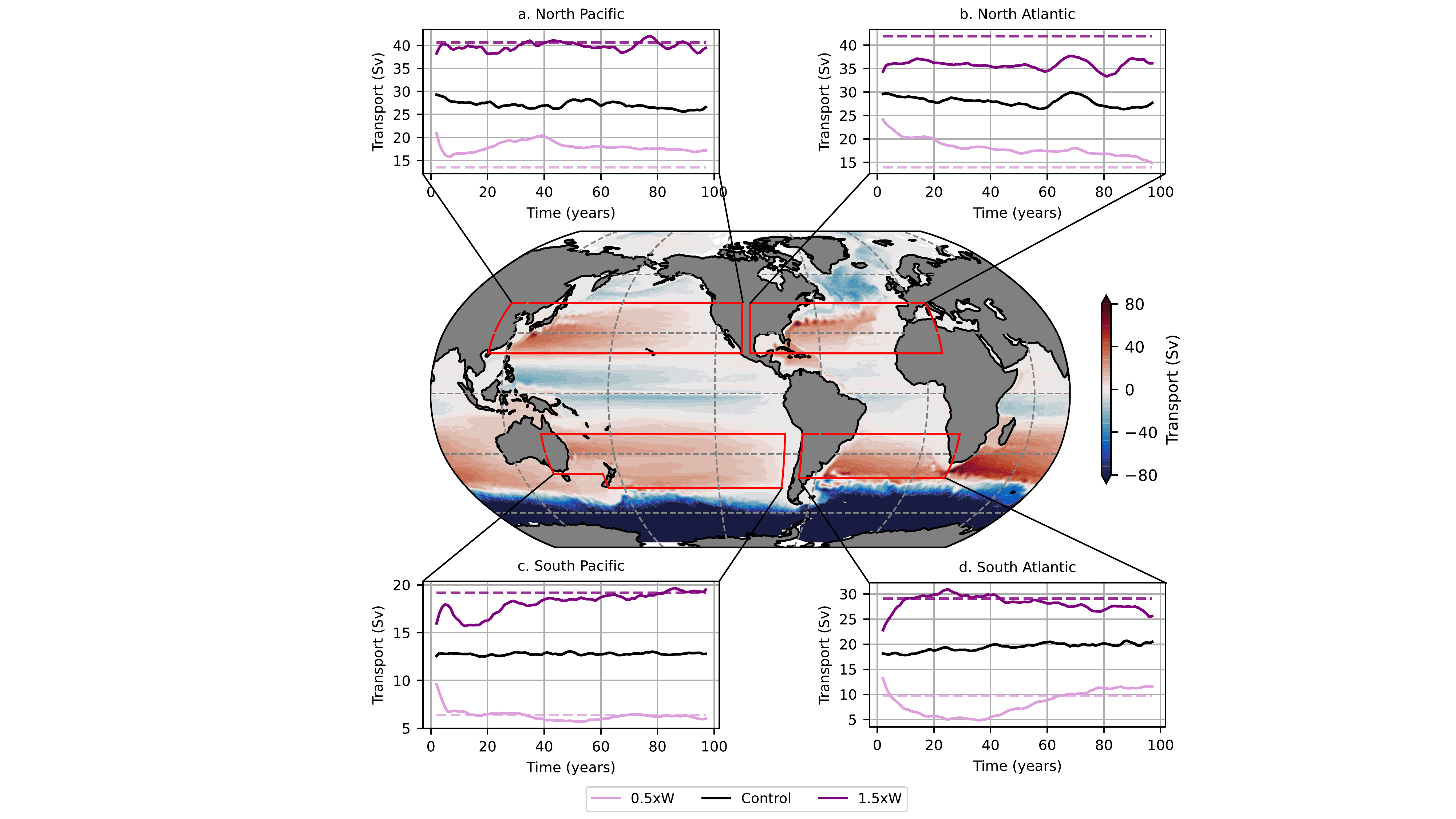}
    \caption{The barotropic streamfunction averaged over the last $15$ years of the flux-forced control experiment. The streamfunction is multiplied with $\mathrm{sign}(f)$. The subpanels show time series of upper-ocean subtropical gyre strength estimated using the isopycnal outcropping method for the control (black), 0.5$\times$W (pink), and 1.5$\times$W (purple) experiments in the (a)~North Pacific, (b)~North Atlantic, (c)~South Pacific, and (d)~South Atlantic. Red boxes indicate the regions we used to estimate the 95$^\text{th}$ percentile subtropical gyre strength. Dashed lines show the gyre transport predictions based on Sverdrup linear scaling, i.e., the time-mean gyre strength in the control multiplied by the wind perturbation factor.}
    \label{fig:wind_driven_gyres}
\end{figure*}

The dashed lines in the time series in Fig.~\ref{fig:wind_driven_gyres} show the expected transport as predicted by the Sverdrup linear scaling of the average gyre transport in the control experiment for the last $100$ years of the simulation. Subtropical gyres follow Sverdrup scaling to a large extent, consistent with the ventilated thermocline theory \citep{Luyten1983}; deviations are observed in the North Atlantic subtropical gyre in both wind perturbation simulations, and in the 0.5$\times$W simulation in the North Pacific subtropical gyre. These deviations from wind stress curl scaling are intriguing and may imply that the gyre is also driven by other mechanisms, including, but not limited to surface buoyancy fluxes; this is discussed in detail in the next two sections.

The gyre strengths adjust quickly to changes in wind forcing (solid curves in the time series in Fig.~\ref{fig:wind_driven_gyres}). This is likely due to a quick adjustment of fast-traveling waves (e.g., large-scale barotropic Rossby waves \citep{Anderson1975} or nearly-barotropic topographic waves \citep{LaCasce2017}). Subsequent smaller-magnitude adjustments in the gyre transports may be attributed to slower wave motions (e.g., baroclinic Rossby waves or surface modes \citep{LaCasce2017}). The time series also show that after the initial response, gyres in the Pacific Ocean are more stable than in the Atlantic Ocean.

\begin{figure}[t]
    \centering
    \includegraphics[width=16.15pc]{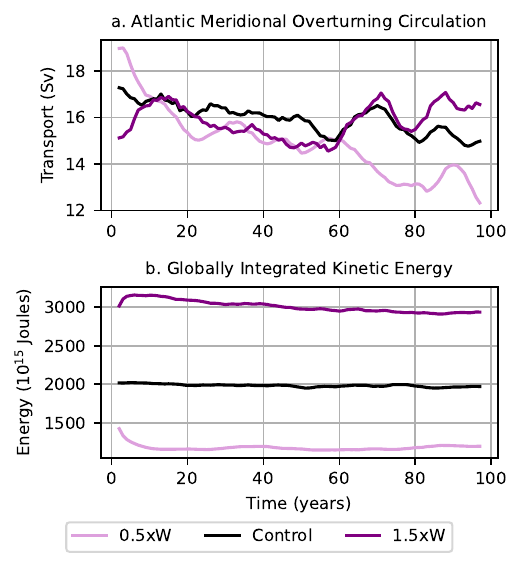}
    \caption{Monthly-mean time-series circulation metrics for the wind perturbation flux-forced simulations: 0.5$\times$W (pink), control (black) and 1.5$\times$W (purple) after a 5-year rolling mean was applied. (a) Atlantic meridional overturning circulation: integrated meridional transport for $\sigma_2 \in [1035.5, 1038.0] \; \mathrm{kg}\,\mathrm{m}^{-3}$ at 26$^{\circ} $N for longitudes between 103$^{\circ} $W and 5$^{\circ}$W.
    (b) Globally integrated kinetic energy.}
    \label{fig:wind_driven_MOC}
\end{figure}

The AMOC (evaluated by integrating horizontal transport between $\sigma_2 \in [1035.5, 1038] \; \mathrm{kg}\,\mathrm{m}^{-3}$ density classes) is only marginally impacted by changes in wind stress. For the first 10 years, the AMOC strength is inversely related to the wind stress magnitude (Fig.~\ref{fig:wind_driven_MOC}a), which is likely linked to a change in northward Gulf Stream barotropic transport (as is also observed by \citet{Hazeleger2006} and \citet{Yang2016}). After the initial transient response, we observe a slight decline in AMOC strength for the 0.5$\times$W experiment compared with the control, but no robust amplification in the 1.5$\times$W experiment, which could be associated with partial eddy compensation in the Southern Ocean \citep{Morrison2013}, or other non-linear feedback in the ocean circulation.


We observe a near-perfect linear scaling of the globally integrated kinetic energy with the wind stress magnitude (Fig.~\ref{fig:wind_driven_MOC}b). Winds supply energy to the ocean through generation of eddies and enhanced circulation, which leads to an increase in the global kinetic energy \citep{Wunsch2004}. We discern that the change in kinetic energy due to wind stress is majorly due to a change in the gyre circulation (Fig.~\ref{fig:wind_driven_gyres}) as well as mesoscale eddies. A complete energy budget calculation is beyond the scope of the study.

\section{Surface buoyancy flux contrast perturbation experiments}

\label{Results_buoy_contrast}
\subsection{Reduced buoyancy flux contrast experiments}

In the previous section, we analyzed the effects of wind stress on the large-scale circulation. In this section, we discuss two sensitivity experiments wherein we reduce the intergyre surface meridional buoyancy difference at the poleward zonal peripheries of subtropical gyres by $7.5$ and $15\,\mathrm{W}\,\mathrm{m}^{-2}$. The surface buoyancy flux contrast perturbation is expected to cause variations in horizontal density gradients: from the thermal wind relation~\eqref{eq:TWR}, these variations will lead to anomalies in the ocean circulation.

\begin{figure*}[h]
    \includegraphics[width=\textwidth]{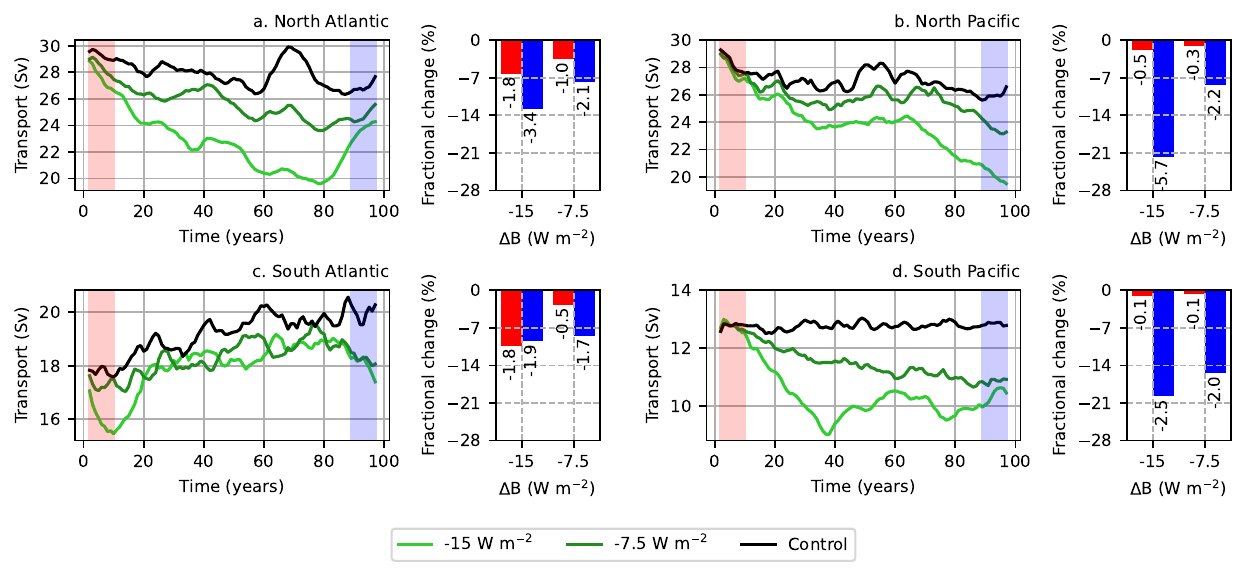}
    \caption{Comparison of subtropical gyre strength for the reduced buoyancy flux contrast experiments. For each gyre, the left panel shows the time series for $-15\;\mathrm{W}\,\mathrm{m}^{-2}$ (dark green), $-7.5\;\mathrm{W}\,\mathrm{m}^{-2}$ (light green), and control (black) simulations, and the right panel shows the fractional change in gyre strength with respect to control for the first 10 (red) and last 10 (blue) years of the simulation. Values on each bar depict the absolute change in gyre strength (in Sv) relative to the control.}
    \label{fig:Neg_buo_gyres_TS}
\end{figure*}

We analyze short ($<1$ decade) and long ($>1$ decade) time responses of the four subtropical gyres to delineate the linear and non-linear behavior of the ocean circulation due to the surface buoyancy flux gradient anomalies. The subtropical gyres, with the exception of the South Pacific gyre, initially reduce compared with the control simulation (Fig.~\ref{fig:Neg_buo_gyres_TS}). This first-decade reduction is approximately linear with respect to the magnitude of the surface buoyancy flux gradient anomaly, as shown by the red bars in Figs.~\ref{fig:Neg_buo_gyres_TS}\mbox{a-c}  for the $-7.5\;\mathrm{W}\,\mathrm{m}^{-2}$ and $-15\;\mathrm{W}\,\mathrm{m}^{-2}$ experiments. The relaxation in gyre strength is consistent with the thermal wind relation~\eqref{eq:TWR}: reduction in buoyancy gradients acts to reduce horizontal flow. The bar graphs alongside the gyre strength time series in Fig.~\ref{fig:Neg_buo_gyres_TS} reveal that the Atlantic subtropical gyres initially react 2 to 4 times more strongly (measured by the percentage change in the gyre transport) to changes in surface heat fluxes than the Pacific subtropical gyres. However, with time, the Pacific gyres display a greater change than the Atlantic gyres. The time series reveal that the Atlantic gyres are more susceptible to a reduction in surface meridional buoyancy forcing contrast than the Pacific Ocean on short timescales.

\begin{figure*}[t]
    \centering
    \vspace{-1em}
    \includegraphics[width=\textwidth]{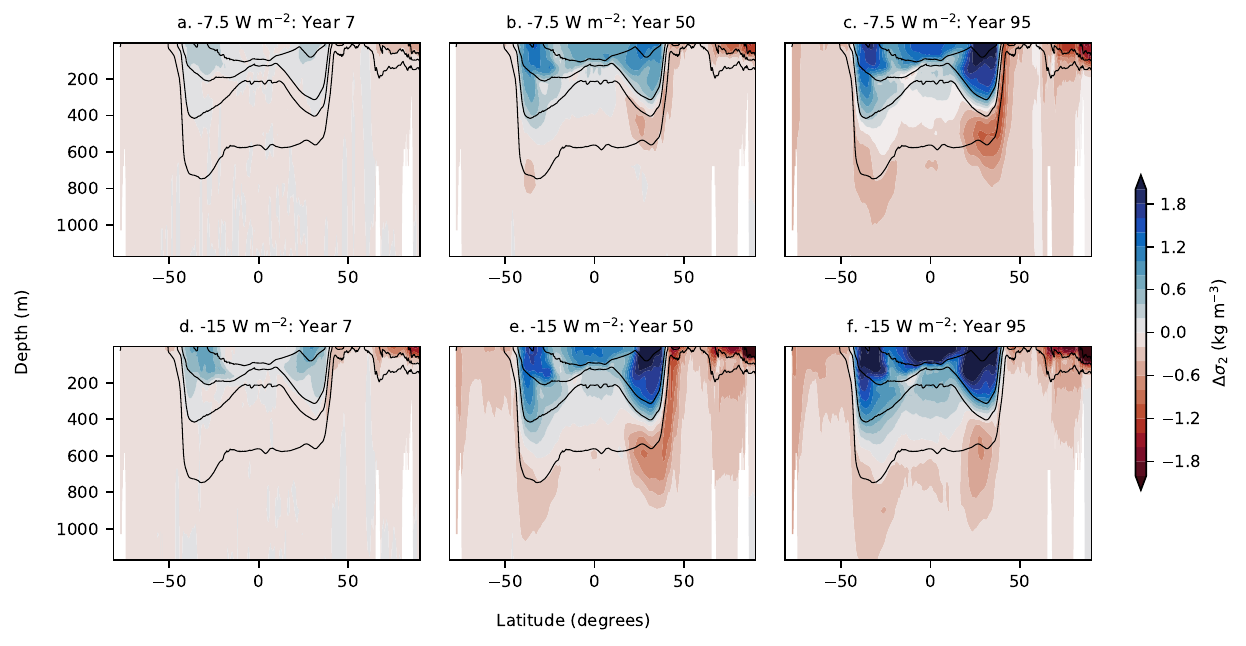}
    \caption{Potential density ($\sigma_2$) anomalies for a longitudinal slice of the upper Atlantic Ocean in the $-7.5\;\mathrm{W}\,\mathrm{m}^{-2}$ (top row) and $-15\;\mathrm{W}\,\mathrm{m}^{-2}$ (bottom row) experiments for year 7 (left column), year 50 (middle column) and year 95 (right column), obtained by averaging between 60$^{\circ}$W and 30$^{\circ}$W for all latitudes. Blue indicates increase in potential density, associated with cooling and/or salinification, whereas red indicates decrease in potential density, associated with heating and/or freshening.}
    \label{fig:neg_buo_PDA_Atlantic}
\end{figure*}

\begin{figure*}[!h]
    \centering
    \vspace{-1em}
    \includegraphics[width=\textwidth]{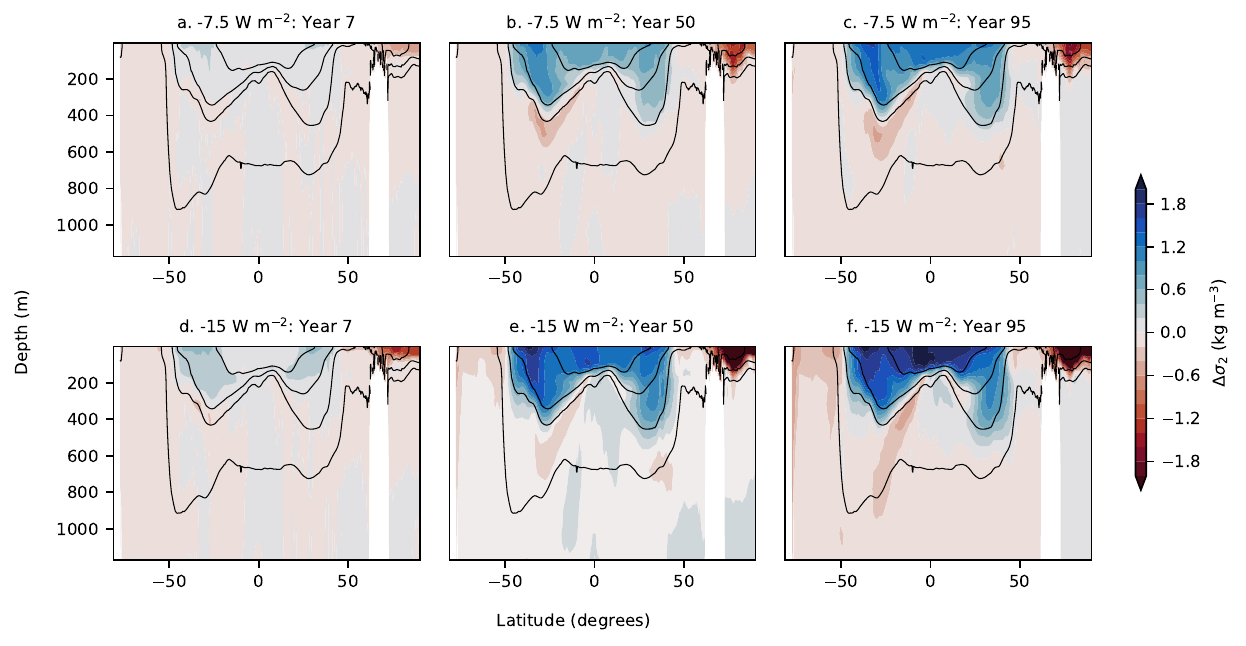}
    \caption{Potential density ($\sigma_2$) anomalies for a longitudinal slice of the upper Pacific Ocean in the $-7.5\;\mathrm{W}\,\mathrm{m}^{-2}$ (top row) and $-15\;\mathrm{W}\,\mathrm{m}^{-2}$ (bottom row) experiments for year 7 (left column), year 50 (middle column) and year 95 (right column), obtained by averaging between 220$^{\circ}$W and 140$^{\circ}$W for all latitudes.}
    \label{fig:neg_buo_PDA_Pacific}
\end{figure*}

Figures~\ref{fig:neg_buo_PDA_Atlantic} and~\ref{fig:neg_buo_PDA_Pacific} highlight spatial and temporal variations in the ocean's density structure due to the applied heat flux anomaly. Focusing on the $-7.5\;\mathrm{W}\,\mathrm{m}^{-2}$ simulation, Figs.~\ref{fig:neg_buo_PDA_Atlantic}a and~\ref{fig:neg_buo_PDA_Pacific}a highlight minor stratification anomalies in the first 7 years of the simulation period, with the subtropical Atlantic Ocean demonstrating slightly larger meridional buoyancy gradient anomalies than the subtropical Pacific Ocean. These gradients are consistent with stronger circulation anomalies (through~\eqref{eq:TWR}) in the two Atlantic subtropical gyres in the initial stages of the simulation. 

In addition to the gyre circulation being linear with respect to the magnitude of the surface buoyancy flux gradient perturbation in the first decade, anomalies in the ocean's buoyancy structure develop linearly with time for the $-7.5\;\mathrm{W}\,\mathrm{m}^{-2}$ simulation. As an example, the potential density latitude-depth transect for the $-7.5\;\mathrm{W}\,\mathrm{m}^{-2}$ simulation at the end of 95 years (Figs.~\ref{fig:neg_buo_PDA_Atlantic}c and~\ref{fig:neg_buo_PDA_Pacific}c) is quite similar to the potential density transect for $-15\;\mathrm{W}\,\mathrm{m}^{-2}$ simulation at the end of 50 years (Figs.~\ref{fig:neg_buo_PDA_Atlantic}e and~\ref{fig:neg_buo_PDA_Pacific}e).


The manifestation of surface buoyancy fluxes on the density structure of the ocean, especially on longer time scales, is not always linear, which may lead to a complex circulatory response. Unlike the $-7.5\;\mathrm{W}\,\mathrm{m}^{-2}$ simulation, Figs.~\ref{fig:neg_buo_PDA_Atlantic}d-f and~\ref{fig:neg_buo_PDA_Pacific}d-f suggest that the anomalies in the Atlantic and Pacific Ocean's density structure in the $-15\;\mathrm{W}\,\mathrm{m}^{-2}$ simulation evolve non-linearly with time in the latter stages of the simulation period due to heat advection by the circulation. Comparing Figs.~\ref{fig:neg_buo_PDA_Atlantic}e and~\ref{fig:neg_buo_PDA_Atlantic}f, we notice an increase in potential density in the upper ocean subtropical region in year 95, which is overshadowed by a relatively stronger (to year 50) potential density increase in the subpolar region. The overall effect is a relative (to year 50) increase in meridional density gradients, and therefore, spin up of the northern region of the subtropical gyre in the $-15\;\mathrm{W}\,\mathrm{m}^{-2}$ simulation by $\approx 18\%$ in the last 20 years of the simulation (Fig.~\ref{fig:Neg_buo_gyres_TS}a). 
In summary, surface buoyancy forcing anomalies alter the density structure of the mixed layer and gradually infiltrate to deeper layers. However, this downward infiltration is continuously modified by the ocean circulation through heat redistribution, leading to a non-linear evolution of the density structure, and hence the gyre circulation on longer timescales.

\begin{figure}[t]
    \centering
    \includegraphics[width=16.15pc]{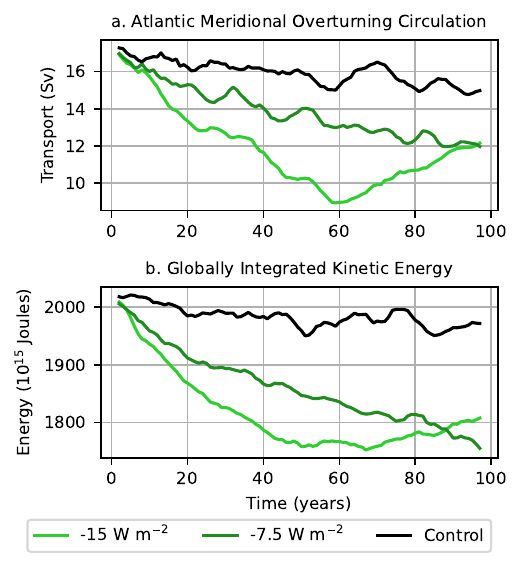}
    \caption{Circulation metrics for the reduced buoyancy flux contrast simulations: $-15\;\mathrm{W}\,\mathrm{m}^{-2}$ (light green), $-7.5\;\mathrm{W}\,\mathrm{m}^{-2}$ (dark green), and control (black) after a 5-year rolling mean was applied. (a) Atlantic meridional overturning circulation: integrated meridional transport for $\sigma_2 \in [1035.5, 1038.0]\; \mathrm{kg}\,\mathrm{m}^{-3}$ at 26$^{\circ} $N for longitudes between 103$^{\circ} $W and 5$^{\circ}$W. (b) Globally integrated kinetic energy.}
    \label{fig:buoyancy_driven_MOC}
\end{figure}

We observe a weakening of the AMOC with time (Fig.~\ref{fig:buoyancy_driven_MOC}a), as a reduction in the surface buoyancy flux gradients strengthens near-surface stratification in the subpolar and polar regions, which suppresses deep water formation in the North Atlantic, a major source of the AMOC \citep{Toggweiler1995, Marshall2012}. The AMOC steadily reduces in the first 60 years, after which it begins to recover in the $-15\;\mathrm{W}\,\mathrm{m}^{-2}$ simulation, as opposed to the $-7.5\;\mathrm{W}\,\mathrm{m}^{-2}$ where it continues to slow down. Moreover, we observe a temporal correlation between the North Atlantic subtropical gyre strength and the AMOC (Figs.~\ref{fig:Neg_buo_gyres_TS}a and~\ref{fig:buoyancy_driven_MOC}a), providing evidence that the two features are interlinked. 

Finally, we briefly discuss the changes in ocean circulation due to surface buoyancy forcing anomalies from an energetics perspective. 
In addition to anomalies observed in the large-scale circulatory features (Figs.~\ref{fig:Neg_buo_gyres_TS} and~\ref{fig:buoyancy_driven_MOC}a), the variations in globally integrated kinetic energy (Fig.~\ref{fig:buoyancy_driven_MOC}b) supplement our understanding that surface buoyancy forcing is an important mechanism in steering the ocean circulation. A reduction in surface buoyancy flux gradients inhibits the production of available potential energy, which in turn reduces the conversion from available potential energy to kinetic energy.

\subsection{Increased buoyancy flux contrast experiments}
In the previous subsection, we considered the short-term and long-term ramifications of reducing surface buoyancy flux gradients on the ocean circulation. A natural follow-up question is: How would the circulation respond to an increase in meridional surface buoyancy flux gradients? Here, we analyze two surface buoyancy perturbation experiments where we increase the meridional surface heat contrast by $+15\;\mathrm{W}\,\mathrm{m}^{-2}$ and $+30\;\mathrm{W}\,\mathrm{m}^{-2}$ at the latitude of western boundary separation for subtropical gyres using the heat flux perturbation map in Fig.~\ref{fig:expt_config}b. 

\begin{figure*}[t]
    \centering
    \includegraphics[width=\textwidth]{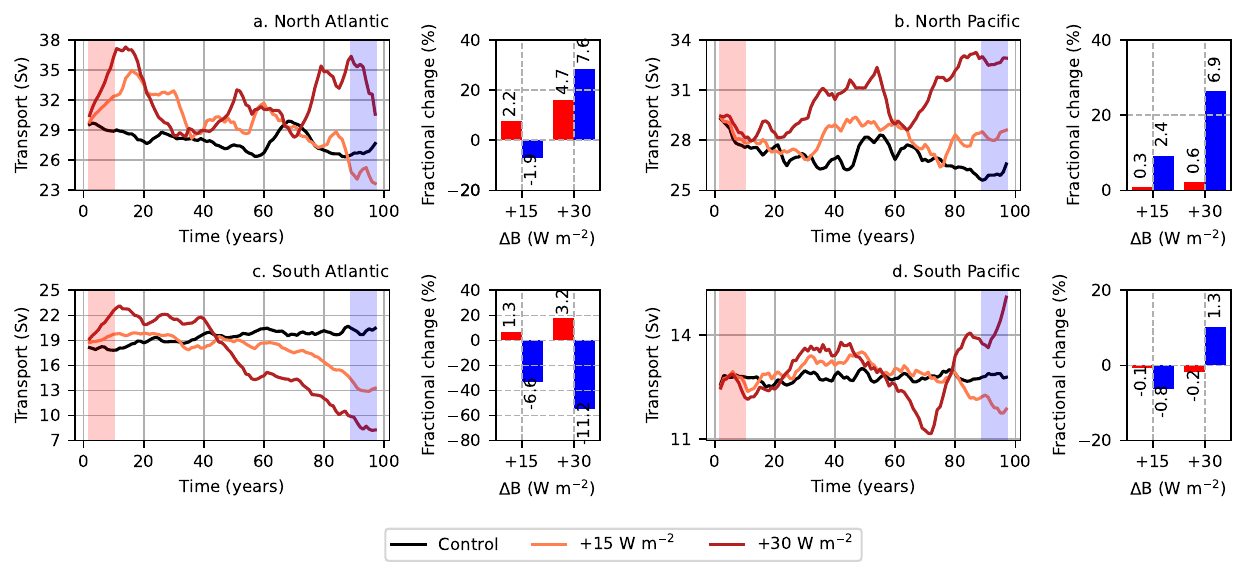}
    \caption{Comparison of subtropical gyre strength for the increased surface buoyancy flux contrast experiments. For each gyre, the left panel shows the time series for control (black), $+15\mathrm{W}\,\mathrm{m}^{-2}$ (orange), and $+30\mathrm{W}\,\mathrm{m}^{-2}$ (red) simulations, and the right panel shows the fractional change in gyre strength with respect to control for the first 10 (red) and last 10 (blue) years of the simulation. Values on each bar depict the absolute change in gyre strength (in Sv) relative to the control.}
    \label{fig:linear_fit_first10}
\end{figure*}

Similar to the reduced buoyancy flux contrast experiments, the anomalies in the Atlantic Ocean in the increased buoyancy flux contrast experiments are induced more quickly than in the Pacific Ocean(compare the red bar graphs in Fig.~\ref{fig:linear_fit_first10}), with the Atlantic gyres in the $+30\;\mathrm{W}\,\mathrm{m}^{-2}$ simulation intensifying by $\sim30\%$ after 15 years. We apply a linear regression model,
\begin{equation}
\label{eq:lin_reg}
    \Delta \psi = m \Delta B,
\end{equation}
to the outputs from the first decade of both the reduced and increased buoyancy flux contrast experiments. In~\eqref{eq:lin_reg}, $\Delta \psi$ is the change in gyre circulation, and $m$ the variation in circulation per unit anomalous surface buoyancy flux contrast $\Delta B$. The regression model predicts a strong linear behavior with reference to the applied surface buoyancy flux contrast for the Atlantic and North Pacific subtropical gyres (inferred from the high $\mathcal{R}^2$ scores in Table~\ref{T3:linear_gyres}). 

Variations in the South Pacific subtropical gyre due to an applied surface buoyancy flux contrast in the first 10 years is negligible (Table~\ref{T3:linear_gyres}). The South Pacific subtropical gyre's erratic response on shorter and longer timescales (Fig.~\ref{fig:linear_fit_first10}d) can be ascribed to the spatial location of the applied surface buoyancy flux gradient perturbation. The gyre encompasses two interconnected sub-gyres with Australia and New Zealand landmasses supporting the two western boundary currents. The anomalous surface buoyancy flux gradient is applied at the southern extent of the subtropical gyre, and as a consequence, the New Zealand sub-gyre is primarily modified. As discussed previously, the isopycnal outcropping method selects the 95$^\text{th}$ percentile of the pseudo-streamfunction in a given basin. Since the New Zealand sub-gyre is weaker than the Australian sub-gyre by a factor of 3 (in the control flux-forced simulation), we do not capture the resulting changes in the South Pacific subtropical gyre time-series. This likely leads to a poor correlation between the gyre strength and the magnitude of surface buoyancy flux contrast.

On longer timescales, the regression model~\eqref{eq:lin_reg} performs poorly, as the relationship between gyre circulation and anomalous surface buoyancy flux contrast becomes non-linear with time. This non-linearity is accompanied by an oscillatory behavior in the gyre strength, as can be observed in the time series in Fig.~\ref{fig:linear_fit_first10}. \citet{Winton1997} argues that these buoyancy-forced oscillations increase in frequency and amplitude with the meridional surface buoyancy flux gradient, consistent with our observations. Although the North Atlantic and North Pacific subtropical gyre strength time series shows high variability (Fig.~\ref{fig:linear_fit_first10}a), the circulation estimates for both $+15\;\mathrm{W}\,\mathrm{m}^{-2}$ and $+30\;\mathrm{W}\,\mathrm{m}^{-2}$ simulations are generally greater than the control. The South Atlantic subtropical gyre strength is enhanced in the first decade of the simulations (Fig.~\ref{fig:linear_fit_first10}c), followed by a plateauing for about 20-25 years. The reduction in the South Atlantic subtropical gyre strength in the last 50 years can be attributed to an inaccurate estimate of the circulation: integrating meridional transport for all isopycnals having $\sigma_2 \le 1035.8 \; \mathrm{kg}\,\mathrm{m}^{-3}$ captures a part of the southward flowing mid-depth overturning circulation. The South Pacific subtropical gyre strength anomalies are minimal in the first decade, and subsequently fluctuate around the control for both $+15\;\mathrm{W}\,\mathrm{m}^{-2}$ and $+30\;\mathrm{W}\,\mathrm{m}^{-2}$ simulations. In conclusion, an oscillatory response is present in all four subtropical gyres, suggesting that there is a complicated feedback between ocean circulation and surface buoyancy forcing. 

\begin{table}[h]
\caption{Linear regression model~\eqref{eq:lin_reg} for the subtropical gyre perturbations over the first $10$ years. $\mathcal{R}^2$ score indicates the extent of gyre variability due to buoyancy forcing that is captured by the linear regression model.
}
\label{T3:linear_gyres}
\begin{center}
\begin{tabular}{ccccrrcrc}
\topline
Gyre basin & $m$ (Sv m$^2 \mathrm{W}^{-1}$) & $\mathcal{R}^2$ score\\
\midline
 North Atlantic & 0.15 &  0.99 \\
 South Atlantic & 0.10 &  0.98 \\
 North Pacific  & 0.02 &  0.92 \\
 South Pacific  & 0.00 & -3.25 \\
\botline
\end{tabular}
\end{center}
\end{table}

\begin{figure*}[t]
    \centering
    \includegraphics[width=38pc]{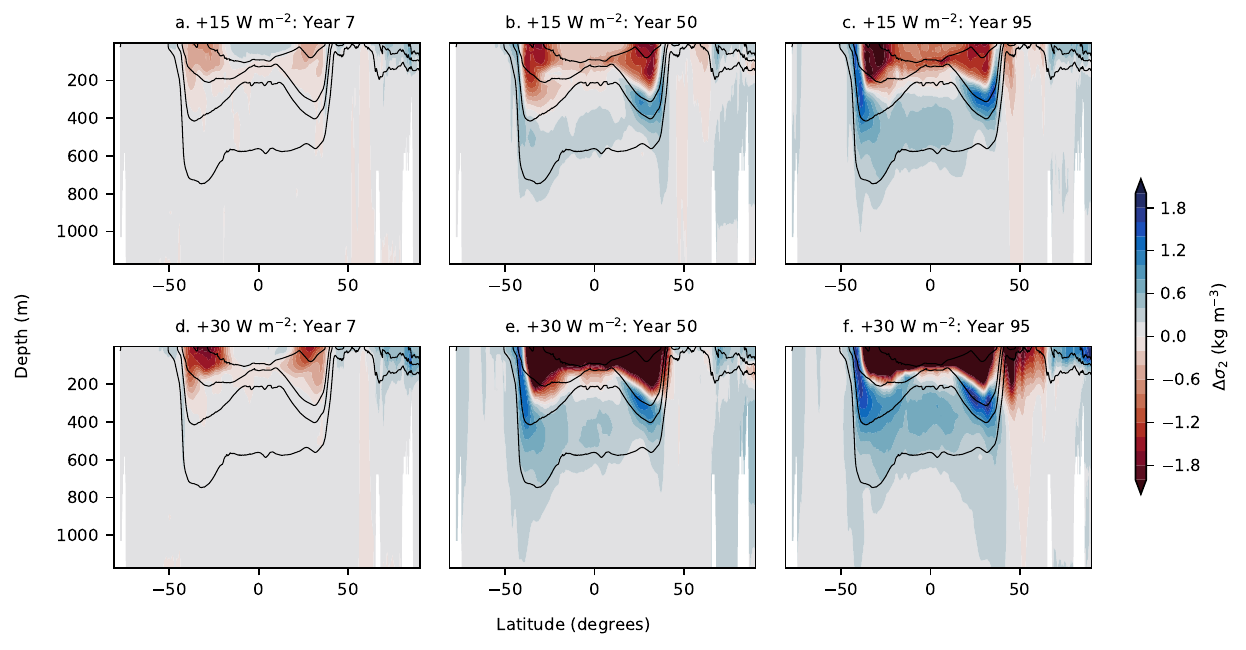}
    \caption{Potential density ($\sigma_2$) anomalies for a longitudinal slice of the upper Atlantic Ocean for $+15\;\mathrm{W}\,\mathrm{m}^{-2}$ (top row) and $+30\;\mathrm{W}\,\mathrm{m}^{-2}$ (bottom row) simulation for year 7 (left column), year 50 (middle column) and year 95 (right column), obtained by averaging between 60$^{\circ}$W and 30$^{\circ}$W for all latitudes.}
    \label{fig:pos_buo_PDA_Atlantic}
\end{figure*}

\begin{figure*}[h!]
    \centering
    \includegraphics[width=38pc]{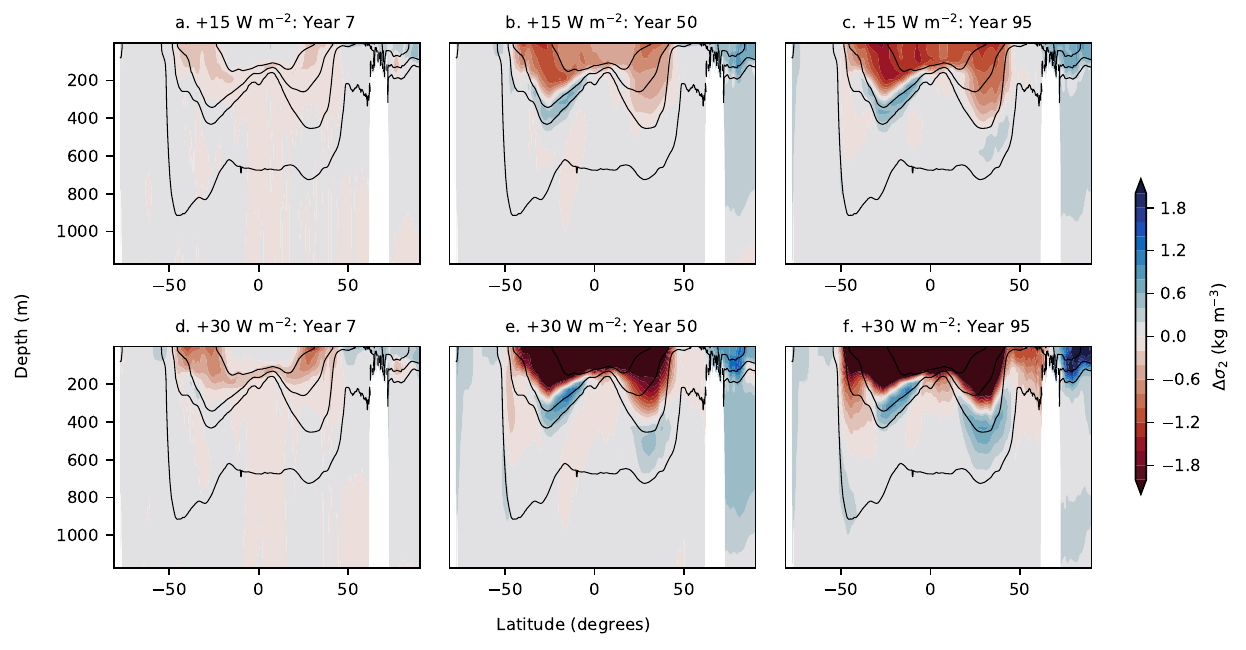}
    \caption{Potential density ($\sigma_2$) anomalies for a longitudinal slice of the upper Pacific Ocean for $+15\;\mathrm{W}\,\mathrm{m}^{-2}$ (top row) and $+30\;\mathrm{W}\,\mathrm{m}^{-2}$ (bottom row) simulation for year 7 (left column), year 50 (middle column) and year 95 (right column), obtained by averaging between 220$^{\circ}$W and 140$^{\circ}$W for all latitudes.}
    \label{fig:pos_buo_PDA_Pacific}
\end{figure*}

In the previous subsection, we observed that the anomalies in the ocean's buoyancy structure due to reduced surface buoyancy flux gradients increase linearly with time. However, this is not true for the increased buoyancy flux contrast experiments (compare the Atlantic basin in Figs.~\ref{fig:pos_buo_PDA_Atlantic}c and ~\ref{fig:pos_buo_PDA_Atlantic}e, and the Pacific basin in Figs.~\ref{fig:pos_buo_PDA_Pacific}c and~\ref{fig:pos_buo_PDA_Pacific}e) where the evolution of the anomalies is more complex especially on longer timescales.

\begin{figure}[t]
    \centering
    \includegraphics[width=16.15pc]{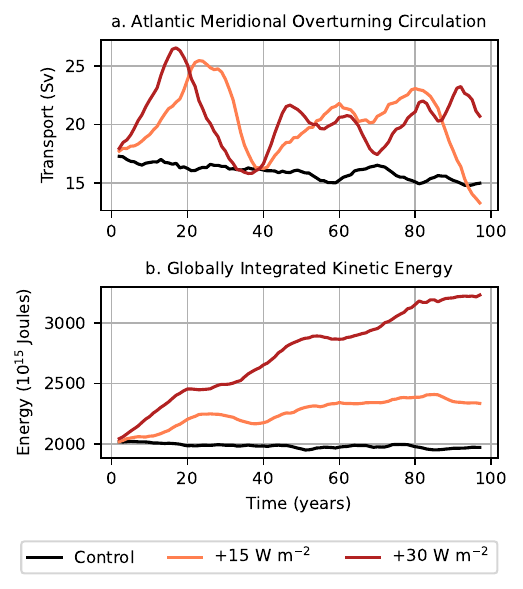}
    \caption{Monthly-mean time-series circulation metrics for the wind perturbation flux-forced simulations: control (black), $+15\;\mathrm{W}\,\mathrm{m}^{-2}$ (orange), and $+30\;\mathrm{W}\,\mathrm{m}^{-2}$ (red) after a 5-year rolling mean was applied. (a) Atlantic meridional overturning circulation: integrated meridional transport for $\sigma_2 \in [1035.5, 1038.0]\; \mathrm{kg}\,\mathrm{m}^{-3}$ at 26$^{\circ} $N for longitudes between 103$^{\circ} $W and 5$^{\circ}$W. (b) Globally integrated kinetic energy.}
    \label{fig:inc_buo_MOC}
\end{figure}

The AMOC in the increased buoyancy flux contrast simulations initially increases, followed by an oscillatory behavior (Fig.~\ref{fig:inc_buo_MOC}a) similar to that observed in the North Atlantic subtropical gyre strength time series in Fig.~\ref{fig:linear_fit_first10}a. 
Increased cooling in subpolar regions stimulates the production of North Atlantic deep water, which causes an acceleration of the mid-depth circulation \citep{Morrison2011}.

An increase in the surface buoyancy flux gradients also promotes the conversion of available potential energy to kinetic energy, which is reflected in Fig.~\ref{fig:inc_buo_MOC}b), consistent with \citet{Hughes2009}. The globally integrated kinetic energy time series reveal an important distinction between wind stress and surface buoyancy forcing: the surface buoyancy flux gradient anomalies alter all aspects of the large-scale ocean circulation (gyres and the AMOC) by varying proportions, unlike wind forcing, which primarily changes the gyre strength by approximately the same factor as the wind stress anomaly. This distinction becomes even more prevalent on longer timescales, as non-linear and non-local feedbacks with the ocean circulation dampens the effect of buoyancy forcing at the surface. 

\section{Uniform warming perturbation experiment}
\label{sec:Global_warming}
In the previous section, we showed that variations in surface heat fluxes, with opposite signed anomalies applied over subpolar and subtropical regions, can produce anomalies in the ocean circulation. In this section we ask; is the same true if the surface heat flux anomalies are globally uniform? 
We expect changes in the ocean circulation due to a spatially uniform surface heat flux due to several processes. Firstly, lateral variations in mixed layer depth imply that buoyancy anomalies induced by the uniform heating will be non-uniform. In addition, changes in circulation continuously alter the buoyancy structure of the ocean through advection. Finally, a spatially varying thermal expansion coefficient implies that a constant surface heat flux manifests as a non-uniform surface buoyancy flux, which can cause non-uniform buoyancy anomalies in the ocean. To understand the combined effects of \emph{(i)}~spatial variations in the mixed layer depth and the thermal expansion coefficient, and \emph{(ii)}~advective feedbacks on the ocean circulation, we analyze a uniform warming experiment, where a globally constant heat flux of $+5\;\mathrm{W}\,\mathrm{m}^{-2}$ is applied at the ocean's surface.

\begin{figure*}[t]
    \centering
    \includegraphics[width=0.85\textwidth]{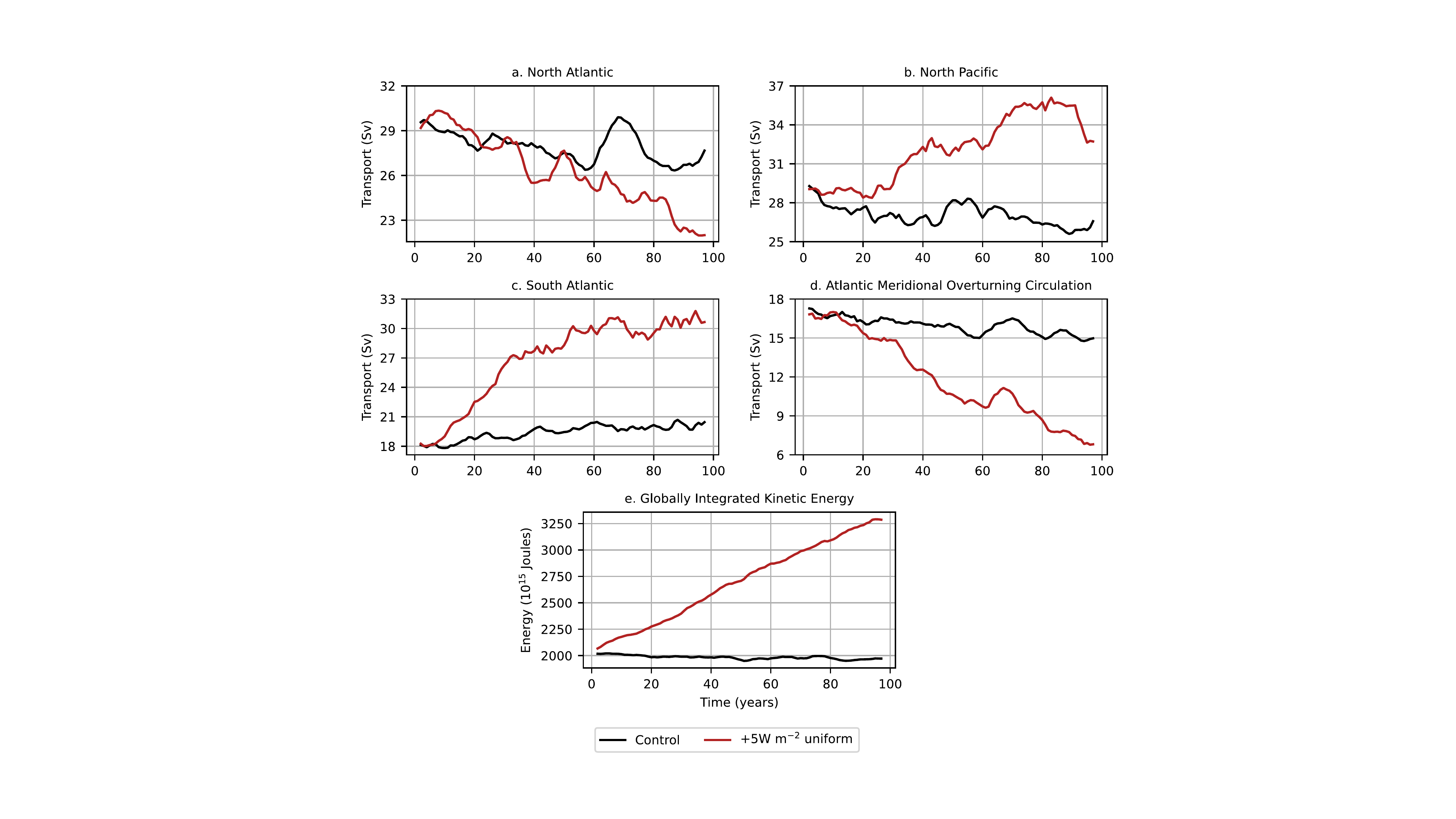}
    \caption{(a)-(c)~Comparison of subtropical gyre strength time series for control (black) and uniform warming (dark red) simulations. (d)~Atlantic meridional overturning circulation: integrated meridional transport for $\sigma_2 \in [1035.5, 1038.0]\; \mathrm{kg}\,\mathrm{m}^{-3}$ at 26$^{\circ} $N for longitudes between 103$^{\circ} $W and 5$^{\circ}$W. (e)~Globally integrated kinetic energy.}
    \label{fig:global_hflux_TS}
\end{figure*}

Changes in the strength of each gyre do occur under the uniform warming perturbation (Fig.~\ref{fig:global_hflux_TS}a-c). Focusing first on the North Pacific basin (Fig.~\ref{fig:global_hflux_TS}b), we observe $\sim$22\% intensification of the subtropical gyre, consistent with the results of \citet{Sakamoto2005} and \citet{Chen2019}. This increase can be attributed to spatial variations in mixed layer depth near the western boundary region.The mixed layer traps the excess heat received from the ocean's surface and distributes only a fraction of this heat to the layer below. Furthermore, a deeper mixed layer has a higher heat capacity due to its ability to store more heat. Deep mixed layers at the western boundary of the North Pacific subtropical gyre moderately shield the region from developing stratification. Conversely, shallower mixed layers to the east of the western boundary lead to more stratification in that region. The spatially uneven growth of stratification in the subtropical gyre strengthens zonal buoyancy gradients near the western boundary region, which in view of the thermal wind relation, intensifies the meridional gyre flow.

We record a strong ($\sim$50\%) intensification of the South Atlantic subtropical gyre (Fig.~\ref{fig:global_hflux_TS}c) due to a similar laterally-varying mixed layer depth as observed in the North Pacific basin, which augments the zonal buoyancy gradients near the western boundary in response to surface heating. The South Pacific gyre anomalies show minor oscillations (with an amplitude of $\sim1.5\,\mathrm{Sv}$) due to uniform surface heating (not shown).

We observe no significant anomalies in the North Atlantic subtropical gyre strength in the first 20 years of the uniform warming simulation (Fig.~\ref{fig:global_hflux_TS}a), followed by a systematic slowdown over the next 75 years. Several modeling studies have reported a correlation between the North Atlantic subtropical gyre strength and the AMOC \citep{Yeager2015, Larson2020}, and we observe a reduction in the AMOC over the same time period (Fig.~\ref{fig:global_hflux_TS}d). This reduction is explained by two processes: \emph{(i)}~in the first 20 years, the Gulf Stream transports slightly larger volume of warm water poleward and \emph{(ii)}~the uniform warming applied at the ocean's surface promotes the generation of lighter waters in the subpolar regions at the ocean's surface. These two processes limit the North Atlantic deep water formation, causing an AMOC slowdown \citep{Lohmann2008, Cheng2013}. 

Finally, the globally integrated kinetic energy increases by almost $50\%$ over the full experiment (Fig.~\ref{fig:global_hflux_TS}e). We can ascribe the resulting kinetic energy increase to mean circulation (for example, the gyres) as well as mesoscale eddies, and is consistent with the energy conversion argument put forth by \citet{Hughes2009} that surface buoyancy forcing could induce kinetic energy in the system through a conversion from available potential energy.

\section{Summary and Discussion}
\label{Conclusions}
We conducted a series of perturbed forcing simulations using a partially eddy-resolving ocean model (at 0.25$^{\circ}$ lateral resolution) to understand the importance of wind stress and surface buoyancy forcing in steering planetary-scale ocean circulation. Our perturbation experiments (listed in Table~\ref{T1:model_setup}) are forced using surface boundary fluxes (and are thus called ``flux-forced simulations'') to separate the contribution of winds and surface buoyancy in driving the circulation, and are classified into three categories: \emph{(i)}~wind perturbation experiments, \emph{(ii)}~surface buoyancy flux contrast perturbations, and \emph{(iii)}~a spatially uniform warming perturbation.

The flux-forced simulations illustrate that both wind stress (Fig.~\ref{fig:wind_driven_gyres}) and surface buoyancy forcing (Figs.~\ref{fig:Neg_buo_gyres_TS},~\ref{fig:linear_fit_first10}, and~\ref{fig:global_hflux_TS}a-c) are crucial in shaping the planetary-scale subtropical gyres. We find that perturbations in surface buoyancy flux gradients modify the ocean's buoyancy structure (Fig.~\ref{fig:neg_buo_PDA_Atlantic},~\ref{fig:neg_buo_PDA_Pacific},~\ref{fig:pos_buo_PDA_Atlantic}, and~\ref{fig:pos_buo_PDA_Pacific}), and thus the circulation through the thermal wind relation~\eqref{eq:TWR}. In addition, spatially uniform surface heat fluxes still induce anomalies in horizontal buoyancy gradients (and hence, the circulation; Fig.~\ref{fig:global_hflux_TS}) due to lateral differences in mixed layer depth and thermal expansion coefficient, and heat advection by the circulation. 

On short timescales ($<1$ decade), the anomalies in horizontal density gradients are proportional to the surface buoyancy flux gradient perturbations (compare panels a~and~d in Figs.~\ref{fig:neg_buo_PDA_Atlantic}, \ref{fig:neg_buo_PDA_Pacific}, \ref{fig:pos_buo_PDA_Atlantic}, and~\ref{fig:pos_buo_PDA_Pacific}). Through the thermal wind relation~\eqref{eq:TWR}, we diagnose a linear relationship ($\mathcal{R}^2 > 0.9$ in Table~\ref{T3:linear_gyres}) between the anomalous gyre circulation and the magnitude of the surface buoyancy flux gradient perturbation on short timescales. Over this period, the Atlantic gyres are observed to be 2-4 times more susceptible to changes in surface buoyancy flux gradients than the Pacific gyres, with as much as a $0.15\,\mathrm{Sv}$ anomaly per $\mathrm{W}\,\mathrm{m}^{-2}$ change in the subtropical/subpolar surface heat flux in the North Atlantic subtropical gyre.

Over time, the induced lateral buoyancy gradients become less proportional to the surface buoyancy flux perturbations (compare panels c~and~f in Figs.~\ref{fig:neg_buo_PDA_Atlantic}, \ref{fig:neg_buo_PDA_Pacific}, \ref{fig:pos_buo_PDA_Atlantic}, and~~\ref{fig:pos_buo_PDA_Pacific}). This divergence from a linear relationship over longer times scales could be attributed to several factors including lateral variations in mixed layer depth and thermal expansion coefficient, and advective feedbacks between the surface buoyancy forcing anomalies and the ocean circulation. For example, in the $-15\,\mathrm{W}\,\mathrm{m}^{-2}$ simulation, this non-linear connection can be observed in the spin-up of the North Atlantic subtropical gyre in the last 20 years (Fig.~\ref{fig:Neg_buo_gyres_TS}a) and a surge in the AMOC in the last 40 years (Fig.~\ref{fig:buoyancy_driven_MOC}a). The ocean circulation in the increased buoyancy flux contrast simulations is more non-linearly related to surface buoyancy flux gradient perturbation than the reduced buoyancy flux contrast simulations. For example, we observe a slowdown in North Atlantic, South Atlantic, and South Pacific gyre strength in the last 40 years of the $+15\,\mathrm{W}\,\mathrm{m}^{-2}$ simulation (Fig.~\ref{fig:linear_fit_first10}) and an oscillatory AMOC in both $+15\,\mathrm{W}\,\mathrm{m}^{-2}$ and $+30\,\mathrm{W}\,\mathrm{m}^{-2}$ simulations (Fig.~\ref{fig:inc_buo_MOC}a).

The flux-forced simulations allowed us to conduct perturbation simulations in which each surface forcing could be altered independently. However, the present study has numerous caveats. Firstly, in reality the wind and buoyancy forcing are strongly coupled. In earlier experiments that used bulk formula for the surface heat fluxes (not shown), we found that decreasing the wind forcing strongly reduced the surface buoyancy fluxes as well due to the reduction in poleward heat transport. Next, the flux-forced simulations are conducted at 0.25$^{\circ}$ resolution and can only partially capture the mesoscale eddies. Moreover, the flux-forced control simulation is not fully equilibrated (see Fig.~\ref{fig:expt_config}c) due to the dynamic frazil formation at high latitudes that continuously adds heat in the polar regions. The frazil formation limits the magnitude of surface buoyancy flux perturbation we can apply in the polar regions. We have partially muted the frazil heat gain in the increased buoyancy flux contrast experiments by adding a globally uniform heat loss. In regions of extreme buoyancy anomalies, the isopycnal outcropping method (section~\ref{Models_methods}b) is prone to capturing other elements of ocean circulation especially in regions of heat gain, such as the deep cell of the AMOC, which may produce erroneous results. Finally, the surface buoyancy flux perturbation experiments have not equilibrated even after 100 years, and hence, should not be misunderstood as the final response.

Our study reinforces recent evidence for a buoyancy-driven component of the ocean gyres \citep{Gjermundsen2018, Hogg2020, Liu2022}. We envisage that a complete theory describing the formation of ocean gyres should incorporate the effects of surface buoyancy forcing, in addition to surface wind stress \citep{Sverdrup1947}. However, the influence of surface buoyancy forcing on gyres depends on the ocean state, and this induces non-linear and non-local feedbacks with the ocean circulation that obscure the formulation of a simple theory. These feedbacks include the role of the mixed layer in capturing excess heat, the non-uniform ingestion of surface heat fluxes by the ocean due to spatially varying thermal expansion coefficient, and the horizontal transport of heat by the circulation, all of which influence the background stratification. 


\acknowledgments
We thank the Consortium for Ocean--Sea Ice Modeling in Australia (\url{http://cosima.org.au}) for helpful discussions and for the development and maintenance of \texttt{cosima-cookbook} package (\url{https://github.com/COSIMA/cosima-cookbook}) and the \texttt{cosima-recipes} repository (\url{https://github.com/COSIMA/cosima-recipes}), both of which are indispensable for our workflow.
D.B.~would like to thank Bishakhdatta Gayen, Aviv Solodoch, Andy Thompson, and Christopher Wolfe for stimulating discussions during early stages of this work.
D.B.~also expresses gratitude to the Computational Modeling Systems group at the Australian Research Council Center for Climate Extremes for their assistance in setting up the flux-forced simulations.
Our analyses were facilitated with the Python packages \texttt{dask} \citep{Rocklin2015} and \texttt{xarray} \citep{Hoyer2017}.
Computational resources were provided by the Australian National Computational Infrastructure at the ANU, which is supported by the Commonwealth Government of Australia.
We acknowledge funding from the Australian Research Council under DECRA Fellowships DE210100004 (R.M.H.) and DE210100749 (N.C.C.).

%

\datastatement
Python code used for generating figures are available at \url{https://github.com/dhruvbhagtani/varying-surface-forcing}. Output to reproduce figures will be available in a Zenodo repository upon acceptance of manuscript. MOM5 source code for flux-forced simulations is available at \url{https://github.com/dhruvbhagtani/MOM5}.

\appendix
\appendixtitle{K-profile ocean surface boundary layer parameterization}
\label{K-profile-appendix}
The ACCESS-OM2 control and flux-forced simulations estimate the mixing layer depth using the K-profile parameterization \citep{Large1994}. The mixing layer depth depends on many factors, such as Langmuir turbulence \citep{Belcher2012}, surface buoyancy forcing \citep{Yoshikawa2015}, wind stress \citep{Grant2011}, and convection \citep{Sohail2020}. The effect of these processes on the mixing layer depth can be encapsulated in the Richardson number, which is the ratio of stratification, $\partial b /\partial z$, and vertical flow shear squared, $|\partial \boldsymbol{u} / \partial z|^2$. The K-profile parameterization uses the bulk Richardson number $\Ri_b$ \citep{Stull1988} defined over a depth $h$:
\begin{equation}
    \label{eq:bulk_rc}
    \Ri_b(h) = \frac{[b(0) - b(-h)] / h}{| \boldsymbol{u}(0) - \boldsymbol{u}(-h)|^2/h^2 + u_{\rm turb}^2/h^2},
\end{equation}
where, e.g., $b(-h) \equiv b(x, y, z=-h, t)$. In~\eqref{eq:bulk_rc}, the numerator is the mean stratification averaged over depth $h$ and in the denominator, $| \boldsymbol{u}(0) - \boldsymbol{u}(-h)| / h$ is the magnitude of the resolved velocity shear averaged over depth $h$ while term $u_{\rm turb} / h$ quantifies the unresolved/turbulent velocity shear. The parameterization determines the mixing layer depth $h$ such that $\Ri_b(h)$ is equal to a critical Richardson number (taken to be $0.3$ in our simulations).

Prior to creating the flux-forced simulations, we conducted wind sensitivity experiments using ACCESS-OM2-025 (not presented here) along with the traditional K-profile parameterization reported in \citet{Large1994}. However, the surface buoyancy forcing was inadvertently modified in these sensitivity experiments through anomalies in the mixing layer depth and ocean circulation. In an attempt to minimize surface buoyancy flux variations in the ACCESS-OM2-025 wind stress sensitivity experiments, we reconstructed the resolved velocity shear term $|\bu(0) - \bu(-h)|/h$ in~\eqref{eq:bulk_rc} in the K-profile parameterization, which was found to primarily cause mixing layer depth anomalies in the sensitivity experiments. We parameterized $|\bu(0) - \bu(-h)|$ as a function of the friction velocity, $u_* = (\boldsymbol{|\tau|} / \rho_0)^{1/2}$ and depth~$h$:
\begin{equation}
\label{parameterisedKPPformula}
    | \bu(0) - \bu(-h) |^2 = (c_a u_*^2 + c_b u_*)  ( 1 - \mathrm{e}^{-c_e h / \sqrt{u_*}} ),
\end{equation}
where $c_a$, $c_b$, and $c_e$ are coefficients determined via multi-variate linear regression. Table~\ref{T2:KPP_coefficients} lists typical ranges of the three parameters for an optimal solution. The parameterization~\eqref{parameterisedKPPformula} performs well in the tropical and subtropical regions. Errors in the polar regions are expected because our parameterization~\eqref{parameterisedKPPformula} does not account for sea ice and marginal ice zone so as to stay consistent with the flux-forced experiments, which could alter resolved velocity shear in these regions.

\begin{table}[h]
   \caption{Parameters for resolved velocity shear obtained using multi-variate linear-regression.}
   \centering
   \label{T2:KPP_coefficients}
   \begin{tabular}{ccccrrcrc} 
   \topline
   Coefficient & Range & Used in our simulations\\ 
   \midline
      $c_a$ & $50-80$ & $80$ \\ 
      $c_b$ & $0.8-1.2$ m s$^{-1}$ & 1 m s$^{-1}$\\
      $c_e$ & $0.009-0.011$ (m s)$^{-1/2}$ & $0.01$ (m s)$^{-1/2}$\\ 
   \botline
\end{tabular}
\end{table} 

The parameterization prevented discrepancies in the mixing layer depth due to alterations in the wind stress, and was implemented in the ACCESS-OM2-025 control experiment (Fig.~\ref{fig:expt_config}c). For consistency with the ACCESS-OM2-025 control simulation, we retained the resolved shear parameterization in the flux-forced simulations. 



\end{document}